\documentclass[11pt, letter]{article}
\pdfoutput=1
\usepackage{amsmath,amssymb,cancel}
\usepackage{stackrel}
\usepackage{graphicx,hyperref,soul}
\usepackage{xcolor}
\usepackage{hyperref, amsfonts, latexsym, bbold, mathbbol}
\usepackage{youngtab}

\usepackage[T1]{fontenc}
\usepackage[utf8]{inputenc}
\usepackage{authblk}

\def\ccr{\nonumber\\} 

\numberwithin{equation}{section}

\begin{document}
\renewcommand\Affilfont{\itshape\footnotesize}

\title{Dressed scalar propagator in a non-abelian background from the worldline formalism}
\author[1]{Naser Ahmadiniaz\thanks{\tiny naser@ifm.umich.mx}}
\author[2]{Fiorenzo Bastianelli\thanks{\tiny bastianelli@bo.infn.it}}
\author[1,3]{Olindo Corradini\thanks{\tiny olindo.corradini@unimore.it}}

\affil[1]{Facultad de Ciencias en F\'isica y Matem\'aticas,
Universidad Aut\'onoma de Chiapas\newline Ciudad Universitaria, Tuxtla Guti\'errez 29050, M\'exico}
\affil[2]{Dipartimento di Fisica ed Astronomia, Universit{\`a} di Bologna\newline and
INFN, Sezione di Bologna, via Irnerio 46, I-40126 Bologna, Italy}
\affil[3]{Dipartimento di Scienze Fisiche, Informatiche e Matematiche, Universit\`a di Modena e Reggio Emilia\newline Via Campi 213/A, I-41125 Modena, Italy}

\maketitle

\abstract{We study  the propagator of a colored scalar particle in the background of a  non-abelian gauge field using the worldline formalism.  It is obtained by considering the open worldline of a scalar particle with extra degrees of freedom needed to take into 
account the color charge of the particle, which we choose to be in the fundamental representation of the 
gauge group. Specializing the external gauge field to be given by a sum of plane waves, i.e. a sum of external gluons, we produce a master formula for the scalar propagator with an arbitrary number of gluons directly attached to the scalar line, akin to similar formulas derived in the literature for the case of the 
scalar particle performing a loop.
 Our worldline description produces at the same time the 
 situation in which the particle has a color charge given by an 
 arbitrarily chosen  symmetric or antisymmetric tensor product of the fundamental.}


\section{Introduction}
\label{sec:Intro}
The worldline representation of effective actions has seen a great deal of activity in the past twenty years, starting with the work of Strassler~\cite{Strassler:1992zr}, who rederived Bern-Kosower master formulas~\cite{Bern:1990cu,Bern:1991aq} directly from point particle path integrals, see ref.~\cite{Schubert:2001he} for a review. Since then many extensions and applications of the worldline formalism have been considered: multi-loop computations~\cite{Schmidt:1994aq}, nonperturbative worldline methods~\cite{Reuter:1996zm, Gies:2005sb,Dunne:2005sx}, the numerical worldline approach to the Casimir effect~\cite{Gies:2003cv},  the worldline formalism in curved spacetime~\cite{Bastianelli:2002fv, Bastianelli:2002qw,Bastianelli:2005vk}, photon-graviton mixing at one loop~\cite{Bastianelli:2004zp}, higher-spin field theory~\cite{Bastianelli:2007pv, Bastianelli:2008nm,Bastianelli:2012bn}, the worldgraph approach to QCD~\cite{Dai:2008bh}, as well as the heat kernel expansion ~\cite{Fliegner:1993wh,Fliegner:1997rk,Bastianelli:2008vh}, 
applications to noncommutative QFT~\cite{Kiem:2001dk,Bonezzi:2012vr}, to the Standard Model physics~\cite{Mansfield:2014vea} and its grand-unified extensions~\cite{Edwards:2014bfa}, just to name a few.

Unlike effective actions, the worldline representation of dressed propagators is still a relatively unexplored land, though a worldline representation for the dressed propagator of a scalar field coupled to electromagnetism (scalar QED) was proposed long ago by Feynman~\cite{Feynman:1951gn}. It consists of a worldline path integral where the coordinate paths have the topology of a line, and come with Dirichlet boundary conditions. 
The problem has then been reconsidered in ref.~\cite{Daikouji:1995dz} along the lines of 
\cite{Strassler:1992zr,Bern:1990cu,Bern:1991aq}. 
More recently, the dressed propagator in a scalar field theory has been studied with worldline methods to address
the summation of ladder and crossed-ladder diagrams and analyze the emergence of bound states~\cite{Bastianelli:2014bfa}.
A full worldline description of dressed QFT propagators would be quite welcome 
in gauge theories as well. It may allow to address several different issues, providing a systematic way of computing scattering amplitudes that could be beneficial both at the perturbative and nonperturbative levels.
At the perturbative level it may improve on the efficiency of perturbative calculations and give
perhaps a better understanding of color/kinematics dualities~\cite{Bern:2008qj}.
 At the nonperturbative level  it might be useful to address the emergence of bound states, or to
  study, for instance, the covariance of the Green's functions under a change in the gauge parameter that gives
rise to the the so-called Landau-Khalatnikov-Fradkin transformations \cite{Fradkin:1955jr,Landau:1955zz}.
 Indeed the worldline formalism has been 
used recently to extend these transformations, originally derived for the scalar propagator,
to an arbitrary $n$-point function in scalar QED~\cite{Ahmadiniaz:2015new}.  The search of computational methods for tree-level amplitudes in gauge theories has been quite an active subject in the past decade. In particular, in scalar QCD, techniques based on recursive relations have been successfully found: see e.g. refs.~\cite{Badger:2005zh, Forde:2005ue, Ferrario:2006np, Boels:2007pj}.

Here we consider the worldline approach to the propagator for a scalar field coupled to an external non-abelian gauge field. Similarly to previous worldline treatments of non-abelian effective actions~\cite{Bastianelli:2013pta, Bastianelli:2015iba}, we obtain the path ordering---needed for the gauge covariance of the worldline path integral---through the quantization of suitable auxiliary variables~\cite{Balachandran:1976ya, Barducci:1976xq,D'Hoker:1995bj}. 
Their usefulness resides in the fact that they allow one to get rid of the explicit path ordering prescription. 
Computationally it is a great advantage, analogous to the replacement of 
the path-ordered spin factors, present in Feynman's original  proposal 
for describing a spin-1/2 particle~\cite{Feynman:1951gn}, with a Grassmann path integral over fermionic 
coordinates \cite{Fradkin:1966zz}.
However, the auxiliary variables thus introduced must be constrained in order to produce an 
irreducible representation (irrep) of the gauge group. One possibility is to couple them to a worldline U(1) 
gauge field, which allows for 
a Chern-Simons term whose coupling constant is chosen to project onto the desired irrep. This is the method
already seen at work in the worldline description of $p$-forms~\cite{Howe:1988ft, Bastianelli:2005vk},
later extended to the treatment of color charges.
Furthermore it allows one to use commuting auxiliary variables which, in absence of the Chern-Simons term, would generate an infinite-dimensional color space. 

Below we show how to employ commuting auxiliary variables with a Chern-Simons term to 
study the propagator of a scalar particle 
in the background of a non-abelian gauge field  by representing it
as a suitable worldline path integral. The non-abelian charge (i.e. color) of the scalar particle can be arbitrary,
though for simplicity we choose it to be in the fundamental representation.
The propagator dressed by the external gauge field is (background) gauge covariant.
One may then specialize the external gauge field to be given by a sum of plane waves
(external gluons)
and perform the path integral. This produces a master formula generating the propagator with an arbitrary 
number of gluons directly attached to it. At fixed number $n$ of gluons,
we call the latter ``partial $n$-gluon scalar propagator'': indeed it consists of 
the scalar propagator with $n$ gluons  directly attached to it, while the gluon self-interactions are excluded.
 As such it is a gauge dependent object. However this dressed propagator is valid 
off-shell, and it can be used as a building block for higher-loop amplitudes.  For example, being valid off-shell, one
 could use it to construct ladder diagrams with gluon rungs, and study the emergence of bound states,
 as done in ref. \cite{Bastianelli:2014bfa}  for the purely scalar case.
 This would still be a gauge-dependent ladder, but in the bound-state studies maintaining
gauge invariance remains a difficult open problem.
One may also try to use it on-shell and study tree-level amplitudes, but then the addition of 
the reducible diagrams with three- and four-gluon vertices
will be essential to achieve gauge invariance.
We do not wish to discuss this final issue here, though one may hope that a kind 
of ``tree replacement rules''---such as those used in similar one-loop master formulas
to generate the missing one-particle-reducible terms
(see for example  ref. \cite{Schubert:2001he})---might work here as well. Alternatively, one may combine the present treatment of color charges with the
worldgraph approach of ref.  \cite{Dai:2008bh}, which is able to generate particle reducible graphs.
We leave this to future analysis.

The main point for us in the present paper is to use and exemplify a novel
representation of the color degree of freedom. The partial $n$-gluon scalar propagator
is the most natural starting point for testing its usefulness in QCD applications.

A final comment on the Chern-Simons coupling: 
fine tuning such coupling allows one to give the scalar particle a color charge
corresponding to any arbitrarily chosen  symmetric tensor product of the fundamental representation,
rather than to the fundamental itself. 
Replacing the commuting variables with anticommuting ones would also give a similar construction, but 
with the non-abelian charge sitting in an arbitrarily chosen antisymmetric tensor product of the fundamental representation,
including the fundamental itself. The choice is again made by selecting an appropriate Chern-Simons coupling.

\section{Tree-level amplitudes in scalar QED}
\label{sec:sQED}
As a warm-up exercise, and in order to fix our notation, we review the computation of tree-level amplitudes in scalar QED from a point particle (worldline) path integral on the line, i.e. with Dirichlet boundary conditions. 

It is well known, since the seminal work of Feynman~\cite{Feynman:1951gn}, that the propagator for a massive charged scalar field coupled to electromagnetism can be obtained from a worldline path integral 
\begin{align}
\langle \phi(x) \bar \phi(x')\rangle_{_A}=\int_{x(0)=x'}^{x(1)=x}\frac{Dx De}{\rm Vol\ Gauge}~e^{-S[x,e;A]}~,
\end{align}
where the particle action in Euclidean signature, is given by (space-time indices are left implicit where not required)
\begin{align}
S[x,e;A]=\int_0^1d\tau \Big( \frac{1}{2e}\dot x^2 +\frac{m^2 e}{2}-iq\dot x\cdot A(x)\Big)~, 
\end{align}
with $e$ being the einbein, i.e. the gauge field for the one-dimensional diffeomorphisms. After the gauge fixing $e=2T$, the path integral over $e$ reduces to a numerical integral over inequivalent constant gauge configurations
labeled by the proper time $T$, i.e.
 \begin{align}
\langle \phi(x) \bar \phi(x')\rangle_{_A}=\int_0^\infty dT~e^{-Tm^2}\int_{x(0)=x'}^{x(1)=x}Dx~e^{-\int_0^1d\tau 
\big(\frac{1}{4T}\dot x^2-iq\dot x\cdot A(x)\big)} ~.
\label{eq:WLA}
\end{align}
In particular, treating the external electromagnetic potential as a perturbation, the expression~\eqref{eq:WLA} can be shown to contain the  sum of an infinite number of tree-level Feynman diagrams with an incoming scalar particle in $x'$, an outgoing scalar particle in $x$, and an arbitrary number of photons.  Specifically, in order to extract the amplitude with $n$ photons, one first writes the potential as a sum of the $n$ photons with polarizations $\varepsilon_l$, and momenta $k_l$,
\begin{align}
A_\mu (x(\tau)) = \sum_{l=1}^n \varepsilon_{\mu,l}~e^{ik_l\cdot x(\tau)}~, 
\end{align} 
and expands the exponential involving the potential, then extracts the amplitude as the term in~\eqref{eq:WLA} that is multilinear in all the different polarizations $\varepsilon_l$'s. The amplitude thus reads
\begin{align}
&{\cal A}(x',x;\varepsilon_1,k_1,\dots \varepsilon_n,k_n)=(iq)^n\int_0^\infty dT~e^{-Tm^2}\prod_{l=1}^n\int_0^1 d\tau_l\ccr
&\times\int_{x(0)=x'}^{x(1)=x}Dx~e^{-\frac{1}{4T}\int_0^1 d\tau \, \dot x^2}~e^{\sum_l(ik_l\cdot x_l+ \varepsilon_l\cdot \dot x_l) }\Big|_{\rm m.l.}~,
\end{align}
where the $\dot x\cdot \varepsilon$ terms have been re-exponentiated  and $x_l:=x(\tau_l)$, whereas ``m.l.'' stands for multilinear. We may now perform the path integral by splitting the generic path $x(\tau)$ into the background, $x_{\rm bg}(\tau) = x' + (x-x')\tau $, satisfying the boundary conditions, and quantum fluctuations $y(\tau)$ with vanishing boundary conditions. We thus get
\begin{align}
&{\cal A}(x',x;\varepsilon_1,k_1,\dots \varepsilon_n,k_n)=(iq)^n\int_0^\infty \frac{dT}{(4\pi T)^{\frac D2}}~e^{-Tm^2-\frac{1}{4T} (x-x')^2}\prod_{l=1}^n\int_0^1 d\tau_l\ccr
&\times e^{\sum_l\big[ik_l\cdot (x'+\tau_l(x-x'))+\varepsilon_l\cdot (x-x')\big]}\Big\langle e^{\sum_l(ik_l\cdot y_l+ \varepsilon_l\cdot \dot y_l) }\Big\rangle\Big|_{\rm m.l.}~,
\label{eq:amp}
\end{align} 
where the expectation value $\big\langle \cdots \big\rangle$ is taken with respect to the free Gaussian path integral $\int_{y(0)=0}^{y(1)=0}Dy~e^{-\frac{1}{4T} \int_0^1 d\tau\, \dot y^2}$. Expression~\eqref{eq:amp} is thus the expectation value of the product of photon vertex operators
\begin{align}
V_A[\varepsilon,k]=e^{ik\cdot x' + \varepsilon\cdot (x-x')}\int_0^1d\tau~e^{\big[i k\cdot(\tau (x-x')+y)+\varepsilon\cdot \dot y \big]} ~.
\label{eq:VA}
\end{align}
From the free path integral one may obtain the worldline propagator
\begin{align}
&\big\langle y^\mu(\tau) y^{\mu'} (\tau') \big\rangle =-2T\delta^{\mu\mu'}\Delta(\tau,\tau')\\
&\Delta(\tau,\tau') =\tau \tau' +\frac12 |\tau -\tau'| -\frac12 (\tau+\tau')\, , \label{eq:Wprop}
\end{align}
and eq.~\eqref{eq:amp} can thus be written as
\begin{align}
&{\cal A}(x',x;\varepsilon_1,k_1,\dots \varepsilon_n,k_n)=(iq)^n\int_0^\infty \frac{dT}{(4\pi T)^{\frac D2}}~e^{-Tm^2-\frac{1}{4T} (x-x')^2}\prod_{l=1}^n\int_0^1 d\tau_l\ccr
&\times \exp\Big\{\sum_{l=1}^n\big[ik_l\cdot (x'+\tau_l(x-x'))+\varepsilon_l\cdot (x-x')\big] 
\ccr
&\hskip1cm+T\sum_{l,l'}\big(k_l\cdot k_{l'}\Delta_{ll'}-i2\varepsilon_l\cdot k_{l'}{}^\bullet\! \Delta_{ll'} -\varepsilon_l\cdot \varepsilon_{l'} {}^\bullet\!\Delta^{\!\bullet}_{ll'}\big) \Big\}\Big|_{\rm m.l.}~,
\label{eq:A}
\end{align}
where $\Delta_{ll'}:= \Delta(\tau_l,\tau_{l'})$, whereas left and right bullets indicate derivatives with respect to $\tau_l$ and $\tau_{l'}$ respectively. Therefore,
\begin{align}
& {}^\bullet\! \Delta_{ll'} = \tau_{l'}-\theta(\tau_{l'}-\tau_l)\\
& {}^\bullet\!\Delta^{\!\bullet}_{ll'} =1-\delta(\tau_{l}-\tau_{l'})~, 
\end{align} 
where $\theta(x)$ is the Heaviside function. Notice that---although products of delta functions appear in the full expansion of the exponential in the expression~\eqref{eq:A}---the  multilinear part of the expansion, that yields the amplitude, does not involve singular, or ill-defined, terms.   

In order to get the amplitude fully in momentum space one may Fourier transform the expression~\eqref{eq:A}, and get
\begin{align}
&{\cal A} (p',p;\varepsilon_1,k_1,\dots \varepsilon_n,k_n) =\int d^Dx \int d^Dx'~e^{i(p\cdot x+p'\cdot x')}  {\cal A}(x',x;\varepsilon_1,k_1,\dots \varepsilon_n,k_n) \ccr
&=\int d^Dx_+ \int d^Dx_-~e^{i(p+p')\cdot x_+ +\frac{i}{2}(p-p')\cdot x_-}  {\cal A}({\textstyle x_+-\frac{x_-}{2}},{\textstyle x_++\frac{x_-}{2}};\varepsilon_1,k_1,\dots \varepsilon_n,k_n)~,
\end{align}  
with $x_-:=x-x'$ and $x_+:=\frac{x+x'}{2}$. The integral over the ``center of mass'' $x_+$ yields the energy-momentum conservation delta-function $(2\pi)^D \delta^{(D)}\big( p+p'+{\textstyle \sum} k_l\big)$ , whereas the integral over the ``distance'' $x_-$ is Gaussian. Hence, after some simple manipulations, the amplitude reduces to
\begin{align}
&\widetilde {\cal A} (p',p;\varepsilon_1,k_1,\dots \varepsilon_n,k_n) =(iq)^n \int_0^\infty dT~e^{-T(m^2+p^2)} \prod_{l=1}^n\int_0^1 d\tau_l\ccr
&\times \exp\Big\{T(p-p')\cdot \sum_{l=1}^n (-k_l \tau_l+i\varepsilon_l) 
+T\sum_{l,l'} \big[k_l\cdot k_{l'}{\mathbb \Delta}_{l-l'}-2i \varepsilon_l\cdot k_{l'}\dot{\mathbb \Delta}_{l-l'} +\varepsilon_l\cdot \varepsilon_{l'}\ddot{\mathbb \Delta}_{l-l'}\big]
\Big\}\Big|_{\rm m.l.}~,
\label{eq:BK-QED}
\end{align}
with ${\mathbb \Delta}_{l-l'} :=\frac12 |\tau_l -\tau_{l'}|$, i.e. the translation-invariant linear part of the worldline propagator~\eqref{eq:Wprop}, a fact that was already noted in ref.~\cite{Daikouji:1995dz}. For simplicity, and for later convenience, in eq.~\eqref{eq:BK-QED} the overall energy-momentum conservation delta function has been stripped off, and the resulting amplitude has been indicated with $\widetilde {\cal A}$.  

\begin{figure}[h]
\begin{center}
\includegraphics[scale=.75]{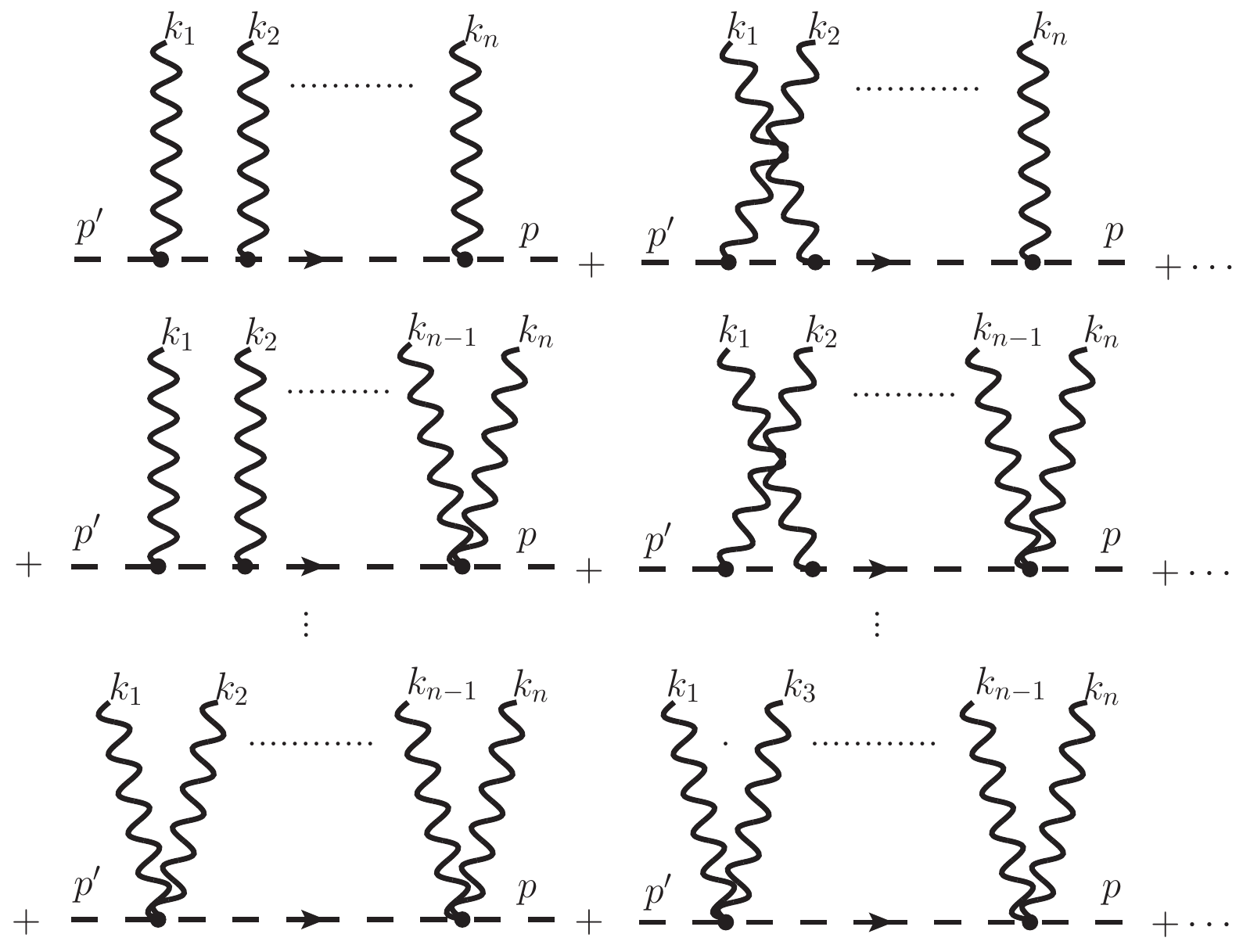}
\end{center}
\caption{\label{fig:n-photon} Diagram structure for the $n$-photon amplitude}
\end{figure}

Equation~\eqref{eq:BK-QED} is the  Bern-Kosower-type formula for the tree-level amplitude with $n$ photons and two scalars, originally worked out in ref.~\cite{Daikouji:1995dz} (see also ref.~\cite{Ahmadiniaz:2015new}).
Note that this amplitude is amputated on the external photon lines, and may be called the $n$-photon scalar 
propagator. It gives the sum of all the QFT Feynman diagrams depicted in Fig.~\ref{fig:n-photon},
which involve linear vertices and seagull vertices, and where the external photon momenta appear in all possible
orderings. This is an advantage of the worldline formalism,
that it combines all orderings into a single expression and that, for scalar QFT, generates the seagull vertices 
within the linear worldline vertex operator.
It is easy to check that, for $n=2$, it reproduces the sum of the three Feynman diagrams responsible for the scalar QED Compton scattering.  Notice that the integrand in eq.~\eqref{eq:BK-QED} is made of a worldline translation-invariant double sum, and a simple sum that is 
not translation-invariant. However it becomes translational invariant 
 on the shell of the scalar lines:
   indeed, imposing the translation $\delta \tau_l =\lambda$ on such sum, and using momentum conservation, it gives $\lambda (p-p')\cdot \sum_l k_l  =\lambda (p'^2-p^2)$, 
which vanishes on-shell.

\section{Dressed scalar propagator in a non-abelian background}
\label{sec:sQCD}
A straightforward way to generalize the previous results to the case of the coupling of a scalar field to a non-abelian gauge field $W_\mu(x) = W_\mu^a(x) T^a$, with $T^a $ belonging to the Lie algebra ${\mathfrak g}$ of a gauge group $G$, is to use a path-ordering prescription to guarantee the gauge covariance of the path integral. Let us thus take a colored scalar field with an index in an irreducible representation of a gauge group, which for definiteness we assume to be $G=SU(N)$. The scalar field propagator in the presence of the non-abelian gauge field can thus be written as
\begin{align}
\langle \phi_\alpha(x) \bar\phi^{\alpha'}(x')\rangle_{_W}= \int_{x(0)=x'}^{x(1)=x}\frac{DxDe}{\rm Vol\ Gauge}~\Big({\cal P}e^{-S[x,e;W]}\Big)_{\alpha}{}^{\alpha'} ~,
\label{eq:WLW}
\end{align}
where the particle action in Euclidean signature is given by
\begin{align}
S[x,e;W]=\int_0^1d\tau \Big( \frac{1}{2e}\dot x^2+\frac{m^2 e}{2} -ig\dot x\cdot W(x)\Big) ~,
\end{align}
and with ${\cal P}$ denoting the path ordering. Although this expression contains implicitly 
the multi-gluon scalar propagator, it may be convenient to investigate an alternative approach that implements the path ordering using some auxiliary variables. For the case of one-loop gluon amplitudes, generated by a scalar field or by a Dirac field, this procedure was proposed in ref.~\cite{D'Hoker:1995bj} and more recently reconsidered in refs.~\cite{Bastianelli:2013pta, Bastianelli:2015iba}. The main advantages of such a procedure are: (i) automatic implementation of path ordering, through the quantization of the auxiliary fields, that allows one to simplify the perturbative calculation. (ii) Simple way to 
extend our formulas to the case of a scalar field in a generic  (anti)symmetric 
tensor product representation of the gauge group. 

For definiteness, let $T^a$ be in the fundamental representation, i.e.  $(T^a)_\alpha{}^{\alpha'}$, $\alpha=1,\dots,N$, then the totally (anti)symmetric tensor $\phi_{\alpha_1\dots\alpha_r}$ transforms in an irreducible (anti)symmetric product of $r$ fundamental representations.
Let us thus consider the addition of complex auxiliary variables, $c_\alpha$ and $\bar c^\alpha$, 
that sit in the fundamental and antifundamental representations of the gauge group, respectively, and that are coupled to the gauge field potential
\begin{align}
\int_0^1d\tau~\Big[ \bar c^\alpha \dot c_\alpha -ig\dot x^\mu W_\mu^a\,  \bar c^\alpha (T^a)_\alpha{}^{\beta} c_{\beta}\Big]-\bar c^\alpha c_\alpha(1)~.
\end{align} 
Note the boundary term  that allows us to set initial conditions on $c$ and final conditions on $\bar c$.
The canonical quantization of such fields gives rise to creation and annihilation operators $\hat c$ 
 and $\hat c^\dagger$ that generate a suitable Fock space. Using a coherent state basis like the one summarized in Appendix~\ref{sec:app}, it yields a scalar wave function in the coordinate $\bar u$,  the left eigenvalue 
 on coherent states of the creation operator $c^\dagger$,
that in turn can be expanded as a $\bar u$-graded sum of fields 
\begin{align}
\phi(x,\bar u) = \phi(x)+\bar u^\alpha \phi_\alpha(x) +\frac{1}{2!} \bar u^{\alpha_1} \bar u^{\alpha_2} \phi_{\alpha_1\alpha_2}(x)+\cdots ~,
\label{eq:Pwave}
\end{align}
so that if the auxiliary variables are (anti)commuting, the tensors $\phi_{\alpha_1\dots \alpha_r}$ sit in an (anti)-symmetrized tensor product of $r$ fundamental representations. In the following we will concentrate on the case of commuting auxiliary variables, as the anticommuting counterpart is just a straightforward generalization. Another necessary ingredient is thus a projector that allows to single out from eq.~\eqref{eq:Pwave} only an irreducible tensor with $r$ indices. This may be achieved by adding a $U(1)$ worldline gauge field to the action, along with a Chern-Simons term. The full locally symmetric action reads 
\begin{align}
S_l[x,c,\bar c;e,a,W]&=\int_0^1d\tau \Big( \frac{1}{2e}\dot x^2+\frac{m^2 e}{2} -ig\dot x^\mu W^a_\mu\, \bar c^\alpha (T^a)_\alpha{}^\beta c_\beta\ccr
&+\bar c^{\alpha}(\partial_\tau+ia)c_\alpha-isa\Big)-\bar c^\alpha c_\alpha(1)~,
\end{align}
where $a(\tau)$ is the $U(1)$ gauge field and $s$ is the Chern-Simons coupling, to be fixed shortly. Due to the presence of the auxiliary fields, the potential is no longer matrix-valued and no explicit path-ordering is thus needed. The propagator  reads
\begin{align}
&\langle \phi(x,\bar u) \bar \phi(x',u')\rangle_{_W}= \int_{x(0)=x',c(0)=u'}^{x(1)=x,\bar c(1)=\bar u}\frac{DxD\bar c DcDeDa}{\rm Vol\ Gauge}~e^{-S_l[x,c,\bar c,e,a;W]}~,
\end{align}
where $u'_\alpha$ and $u_\alpha$ represent the initial and final color states of the scalar field. Thus, if the scalar field sits in the rank-$r$ totally symmetric representation, the above field operators are (apart from numerical factors) given by
\begin{align}
&\phi(x,\bar u) \sim \phi_{\alpha_1\cdots \alpha_r}(x) \bar u^{\alpha_1}\cdots \bar u^{\alpha_r}\,,\quad \bar \phi(x',u') \sim \bar \phi^{\alpha_1\cdots \alpha_r}(x') u'_{\alpha_1}\cdots u'_{\alpha_r}~.
\end{align}

   Upon gauge-fixing the einbein $e$ can be set to a modulus $2T$ as above, whereas the abelian gauge field can be set to a constant angle modulus $\theta \in [0,2\pi)$. The gauge-fixed path integral thus reads
\begin{align}
&\langle \phi(x,\bar u) \bar \phi(x',u')\rangle_{_W}\nonumber \\&= \int_0^\infty dT~e^{-Tm^2} \int_0^{2\pi}\frac{d\theta}{2\pi}~e^{is\theta}\int_{x(0)=x'}^{x(1)=x} Dx\int_{c(0)=u'}^{\bar c(1)=\bar u}D\bar c Dc~e^{-S[x,c,\bar c,2T,\theta;W]}~,
\label{eq:WLgf}
\end{align}  
with 
\begin{align}
S[x,c,\bar c;2T,\theta,W]&=\int_0^1d\tau \Big( \frac{1}{4T}\dot x^2 -ig\dot x^\mu W^a_\mu\, \bar c^\alpha (T^a)_\alpha{}^\beta c_\beta
+\bar c^{\alpha}(\partial_\tau+i\theta)c_\alpha\Big)-\bar c^\alpha c_\alpha(1)~,
\end{align}
where the constant part $-Tm^2 +is\theta$ was stripped off. Classically the equation of motion for $a$ corresponds to the constraint $\bar c^\alpha c_\alpha-s=0$. In the canonical quantization approach this turns into an operatorial constraint to be imposed {\em \`a la Dirac-Gupta-Bleuler} on the wave function.
 As shown in the Appendix~ \ref{sec:app} we choose to identify the path integral expression $ c^\dagger{}^\alpha c_\alpha$ as the one representing the normal-ordered hamiltonian $ \hat c^\dagger{}^\alpha \hat c_\alpha $. Therefore the quantum constraint reads
 \begin{align}
\left(\bar c^\alpha \frac{\partial}{\partial \bar c^\alpha} -s\right)\phi(x,\bar c) = 0~,
\end{align}     
and thus, in order to single out the tensor with $r$ indices, we identify $s=r$, and set the path integral normalization to (see eq.~\eqref{eq:pathN})
 \begin{align}
\int_{c(0)=u'}^{\bar c(1)=\bar u} D\bar c Dc ~e^{-\int_0^1d\tau\, \bar c^\alpha\big( \partial_\tau+i\theta \big) c_\alpha +\bar c^\alpha c_\alpha(1)}=e^{e^{-i\theta}\bar u \cdot u'}~,
\label{eq:pathN'}
\end{align}
with $\bar u\cdot u' = \bar u^\alpha u'_\alpha$. In the following we find it convenient to twist the auxiliary fields in order to absorb the $\theta$ term, i.e. $c(\tau) \to c(\tau) e^{-i\theta\tau}$ and $\bar c(\tau) \to \bar c(\tau) e^{i\theta\tau}$, so that the previous expression gets replaced by
\begin{align}
\int_{c(0)=u'}^{\bar c(1)=e^{-i\theta}\bar u} D\bar c Dc ~e^{-\int_0^1d\tau\, \bar c^\alpha\dot c_\alpha +\bar c^\alpha c_\alpha(1)}=e^{e^{-i\theta}\bar u \cdot u'}~,
\label{eq:pathN''}
\end{align} 
and~\eqref{eq:WLgf} reduces to
\begin{align}
\langle \phi(x,\bar u) \bar \phi(x',u')\rangle_{_W}
&= \int_0^\infty dT~e^{-Tm^2} \int_0^{2\pi}\frac{d\theta}{2\pi}~e^{ir\theta}\int_{x(0)=x'}^{x(1)=x} Dx\int_{c(0)=u'}^{\bar c(1)=e^{-i\theta}\bar u}D\bar c Dc \nonumber \\& \times e^{-\int_0^1d\tau \big( \frac{1}{4T}\dot x^2 -ig\dot x^\mu W^a_\mu\, \bar c^\alpha (T^a)_\alpha{}^\beta c_\beta
+\bar c^{\alpha}\dot c_\alpha\big)+\bar c^\alpha c_\alpha(1)}~.
\label{eq:WLgff}
\end{align}
Splitting as usual the fields into backgrounds and fluctuations
\begin{align}
&c_\alpha(\tau) = u'_\alpha + \kappa_\alpha(\tau)\,, \quad \kappa_\alpha(0)=0\\
&\bar c^\alpha(\tau) =e^{-i\theta}\bar u^\alpha +\bar \kappa^\alpha(\tau)\,,\quad \bar\kappa^\alpha(1) =0~,
\end{align}
the kinetic action for the fluctuations $\int_0^1 d\tau \bar\kappa^\alpha \dot \kappa_\alpha$ can be inverted to give the propagator
\begin{align}
\langle \kappa_\alpha(\tau) \bar \kappa^\beta(\sigma) \rangle = \delta_\alpha^\beta\, \theta(\tau-\sigma)~.
\label{eq:Kprop}
\end{align}

Eq. \eqref{eq:WLgff} gives the final form of 
our worldline representation for the scalar propagator dressed by a non-abelian gauge field.
It is gauge-covariant, and we discuss the related Ward identities at the end of next section. 
For an arbitrary external gauge field configuration one is not able to perform the path integral exactly.\
On the other hand, one may use it for suitable perturbative expansion, or to get exact results   
for specific external field configurations.

As our main application we specialize  the external gauge potential to be given by a sum of plane waves.
This allows for an exact path integral calculation, and 
produces a master formula for 
the scalar propagator with an arbitrary number $n$ of external (amputated) gluons lines directly 
attached to it, generating what we have called the ``partial $n$-gluon scalar propagator''.
As already discussed, the final result will not be gauge invariant, but it may be used as a
starting building block to get eventually gauge invariant amplitudes. Also note that our master formula
is similar to the one-loop Bern-Kosower-type of master formulas used in scalar QCD, that contain
all one-particle irreducible diagrams with a scalar loop and an arbitrary number of external 
(amputated) gluon lines, the only difference being that instead of a loop we have an open line for the scalar particle.

Thus, in order to extract
our partial $n$-gluon scalar propagator 
 from the gauge-fixed worldline path integral~\eqref{eq:WLgff}  we write the potential as a sum of $n$ gluons with polarizations $\varepsilon_l$, momenta $k_l$ and colors $a_l$, i.e.
\begin{align}
W_\mu(x) = \sum_{l=1}^n\varepsilon_\mu(k_l) T^{a_l}\, e^{ik_l\cdot x}~, 
\end{align}     
and this allows us to read off the gluon vertex operator from the interaction term, namely
\begin{align}
V_W[\varepsilon,k,a] := e^{ik\cdot x' + \varepsilon\cdot (x-x')}\int_0^1d\tau~e^{\big[i k\cdot(\tau (x-x')+q)+\varepsilon\cdot \dot q \big]}\big(e^{-i\theta} \bar u^\alpha+\bar \kappa^\alpha\big) (T^a)_\alpha{}^\beta \big( u_\beta+ \kappa_\beta\big)\Big\vert_{\rm lin.\, \varepsilon}~,
\label{eq:VW} 
\end{align}
which differs from the photon vertex operator of eq.~\eqref{eq:VA} in the presence of the color factor, given by the color generator $T^a$ contracted with the auxiliary fields. The $n$-gluon two-scalar term  reads
\begin{align}
&{\cal A}_W(p,u;p',u';\varepsilon_1,k_1,a_1,\dots,\varepsilon_n,k_n,a_n) = (ig)^n\int d^Dx\, d^Dx'~e^{i(p\cdot x+p'\cdot x')}
\ccr &\times\int_0^\infty\frac{dT}{(4\pi T)^{\frac{D}{2}}}~e^{-Tm^2-\frac{1}{4T}(x-x')^2}\int_0^{2\pi} \frac{d\theta}{2\pi}~e^{ir\theta+e^{-i\theta}\bar u\cdot u'}\prod_{l=1}^n\int_0^1d\tau_l
\ccr & \times\Big\langle V_W[k_1,\varepsilon_1,a_1]\cdots V_W[k_n,\varepsilon_n,a_n]\Big\rangle~,
\label{eq:Wamp}
\end{align}
where the average is computed with the propagators~\eqref{eq:Wprop} and~\eqref{eq:Kprop}. Due to the fact that the color part of the gluon vertex operator factors out, the integrand of the non-color part of the expression~\eqref{eq:Wamp} is  identical to its photon counterpart given in ref.~\eqref{eq:BK-QED}. We thus obtain the following final result for the tree-level 
partial $n$-gluon scalar propagator
\begin{align}
&\widetilde{\cal A}_W(p,u;p',u';\varepsilon_1,k_1,a_1,\dots,\varepsilon_n,k_n,a_n) = (ig)^n
\int_0^\infty dT~e^{-T(m^2+p^2)}\prod_{l=1}^n\int_0^1d\tau_l\ccr
&\times \exp\Big\{T(p-p')\cdot \sum_l (-k_l \tau_l+i\varepsilon_l) 
+T\sum_{l,l'} \big[k_l\cdot k_{l'}{\mathbb \Delta}_{l-l'}-2i \varepsilon_l\cdot k_{l'}\dot{\mathbb \Delta}_{l-l'} +\varepsilon_l\cdot \varepsilon_{l'}\ddot{\mathbb \Delta}_{l-l'}\big]
\Big\}\Big|_{\rm m.l.}\ccr
&\times \int_0^{2\pi} \frac{d\theta}{2\pi}~e^{ir\theta+e^{-i\theta}\bar u\cdot u'}
\Big\langle \big(e^{-i\theta} \bar u^{\alpha_1}+\bar \kappa^{\alpha_1}(\tau_1)\big)(T^{a_1})_{\alpha_1}{}^{\beta_1}\big( u_{\beta_1}+ \kappa_{\beta_1}(\tau_1)\big)\cdots\ccr & \hskip4cm\cdots\big(e^{-i\theta} \bar u^{\alpha_n}+\bar \kappa^{\alpha_n}(\tau_n)\big)(T^{a_n})_{\alpha_n}{}^{\beta_n} \big( u_{\beta_n}+ \kappa_{\beta_n}(\tau_n)\big)\Big\rangle~,
\label{eq:BK-QCD}
\end{align}
whose corresponding QFT Feynman diagrams have the same structure as those depicted in Fig.~\ref{fig:n-photon} for the $n$-photon amplitude. Let us underline that, since no three- and four-gluon QCD vertices are involved in our worldline computation,  expression~\eqref{eq:BK-QCD} only represents the gluon-irreducible part of the full $n$-gluon two-scalar  amplitude, i.e. the part of the amplitude made of Feynman diagrams that cannot be parted by cutting a gluon internal line.     

Before studying some special case let us consider the $\theta$ integrals present in the previous Bern-Kosower-type formula. Let us define
\begin{align}
F(k,\bar u\cdot u'):= \int_0^{2\pi} \frac{d\theta}{2\pi}~e^{ik\theta +e^{-i\theta} \bar u\cdot u'}~,
\end{align} 
with $k\in {\mathbb Z}$. We can solve it by transforming it into a clockwise contour integral over the unit circle: let us define $z:=e^{-i\theta}$. Hence,
\begin{align}
F(k,\bar u\cdot u')= \oint \frac{dz}{-2\pi i z}~\frac1{z^k}\, e^{\bar u\cdot u' z} =\left \{ 
\begin{array}{ll}
\frac{1}{k!}(\bar u\cdot u')^k\,,\quad  & k\geq 0\\[1mm]
0\,, & k<0
\end{array}
\right.~.
\end{align} 
Notice that, although the function $e^{\bar u\cdot u' z}$ presents an essential singularity at $z\to\infty$, it is a holomorphic function in any bounded domain, and it can be written as a power series. Thus, the contour integral over the unit circle picks out  the order $k$ coefficient of such power series.  
  
Let us now single out a few examples in order to further clarify the results obtained. 

\subsection{Free scalar propagator}
The free scalar propagator, for a field in a totally symmetric rank-$r$ representation, can obviously be obtained from the above formalism by considering zero external gluons ($n=0$). This will help us to fix an overall prefactor. For $n=0$, eq.~\eqref{eq:BK-QCD} reduces to  
\begin{align}
\widetilde{\cal A}_W(p,u;-p,u') = \frac{1}{p^2+m^2}\, F(r,\bar u\cdot u') = \frac{\bar u^{\alpha_1}\cdots \bar u^{\alpha_r} u'_{\beta_1}\cdots u'_{\beta_r}}{r!}\, \frac{\delta^{\beta_1\cdots \beta_r}_{\alpha_1\cdots \alpha_r}}{p^2+m^2}~,
\label{eq:prop-pol}
\end{align}  
where $\delta^{\beta_1\cdots \beta_r}_{\alpha_1\cdots \alpha_r}:= \delta^{(\beta_1}_{\alpha_1}\cdots \delta^{\beta_r)}_{\alpha_r}$ is the identity in the totally symmetric rank-$r$ representation. Equation~\eqref{eq:prop-pol} corresponds to the free propagator for the polarized scalar field, and 
one may obtain the free propagator for the unpolarized scalar field by stripping off the prefactor $\frac{\bar u^{\alpha_1}\cdots \bar u^{\alpha_r} u'_{\beta_1}\cdots u'_{\beta_r}}{r!}$ from~\eqref{eq:prop-pol}. In fact note that 
\begin{align}
\frac{\bar u^{\alpha_1}\cdots \bar u^{\alpha_r} u'_{\beta_1}\cdots u'_{\beta_r}}{r!}\, {\delta^{\beta_1\cdots \beta_r}_{\alpha_1\cdots \alpha_r}} = {\mathbb P}_r \langle \bar u\vert u'\rangle = {\mathbb P}_r \, e^{\bar u\cdot u'}~,
\end{align}
where ${\mathbb P}_r$ is the projector on the above irrep---basically it picks up the order $r$ term in the Taylor expansion of the exponent. Furthermore, using the identity
\begin{align}
\frac{1}{r!}\int \prod_{\alpha=1}^N \frac{d\bar u^\alpha u_\alpha}{2\pi i}e^{-\bar u\cdot u} u_{\alpha_1}\cdots u_{\alpha_r} \bar u^{\beta_1}\cdots \bar u^{\beta_r} &= 
\frac{1}{r!} \frac{\partial^r}{\partial \bar v^{\alpha_1}\cdots \partial \bar v^{\alpha_r}} \frac{\partial^r}{\partial v_{\beta_1}\cdots \partial v_{\beta_r}}\, e^{\bar v\cdot v}\Big\vert_{v=\bar v=0}\nonumber \\& = \delta^{\beta_1\cdots \beta_r}_{\alpha_1\cdots \alpha_r}~,
\label{eq:compose}
\end{align}
 one can easily prove the following composition rule for the free propagator
\begin{align}
&\int d^Dx' \int \prod_{\alpha=1}^N \frac{d\bar u'^\alpha u'_\alpha}{2\pi i}e^{-\bar u'\cdot u'} \langle x,\bar u\vert \frac{{\mathbb P}_r}{\hat p^2+m^2} \vert x',u'\rangle \langle x',\bar u'\vert \frac{{\mathbb P}_r}{\hat p^2+m^2} \vert x'',u''\rangle \ccr
&= \langle x,\bar u\vert \frac{{\mathbb P}_r}{(\hat p^2+m^2)^2} \vert x'',u''\rangle~.
\end{align}
In eq.~\eqref{eq:compose} we used the sources $v_\alpha$ and $\bar v^\alpha$ to insert the terms  $u_{\alpha_1}\cdots u_{\alpha_r} \bar u^{\beta_1}\cdots \bar u^{\beta_r}$ into the Gaussian integral.

Of course the prefactor $\frac{\bar u^{\alpha_1}\cdots \bar u^{\alpha_r} u'_{\beta_1}\cdots u'_{\beta_r}}{r!}$ is universal, i.e. it only depends on the two external scalar lines and not on the gluon vertices. Therefore the recipe to get the
generic $n$-gluon two-scalar term from eq.~\eqref{eq:BK-QCD}, is to strip off such color prefactors and truncate (i.e. multiply by the inverse free propagators) the external scalar lines.

\subsection{Partial $n$-gluon scalar propagator}
For the simplest case of one gluon, expression~\eqref{eq:BK-QCD} reproduces the Feynman 
vertex linear in the gluon,  whose color part is given by the expression
\begin{align}
&F^a(\bar u,u')=\oint \frac{dz}{-2\pi i z}~\frac1{z^r}\, e^{\bar u\cdot u' z} \Big\langle \left({\bar u^{\alpha}}{z}+\bar \kappa^{\alpha}(\tau)\right)(T^{a})_{\alpha}{}^{\beta}\big( u_{\beta}+ \kappa_{\beta}(\tau)\big)\Big\rangle \ccr
& =\oint \frac{dz}{-2\pi i z}~\frac1{z^{r-1}}\, e^{\bar u\cdot u' z}\, \bar u T^{a_1} u' =\frac{1}{(r-1)!} 
(\bar u\cdot u')^{r-1} \bar u T^{a} u' \ccr
&= \frac{1}{r!}\bar u^{\alpha_1}\cdots \bar u^{\alpha_r}  (T_{(r)}^{a})_{\alpha_1\dots\alpha_r}^{\beta_1\cdots \beta_r}u'_{\beta_1}\cdots u'_{\beta_r}~,
\end{align}
where
\begin{align}
(T_{(r)}^{a})_{\alpha_1\dots\alpha_r}^{\beta_1\cdots \beta_r} =r (T^a)_{(\alpha_1}{}^{(\beta_1}\delta_{\alpha_2}^{\beta_2}\cdots \delta_{\alpha_r)}^{\beta_r)}\,,\quad \quad\quad
\underbrace{\, \yng(5) \, }_{r}~,
\end{align}
is a generator of $SU(N)$ for the rank-$r$ totally symmetric representation. Above we used the fact that $\langle \kappa_\beta(\tau) \bar \kappa^\alpha(\tau)\rangle =\frac12 \delta_\beta^\alpha$ and that $T^a$ is traceless. Thus, 
 it reads
\begin{align}
&\widetilde{\cal A}_W(p,u;p',u';\varepsilon,k,a) = ig \int_0^\infty dT~e^{-T(m^2+p^2)}\int_0^1d\tau ~Te^{-T(k^2+2p\cdot k \tau)}\ccr
&\times i\varepsilon\cdot (p-p')~\frac{1}{r!}\bar u^{\alpha_1}\cdots \bar u^{\alpha_r}  (T_{(r)}^{a})_{\alpha_1\dots\alpha_r}{}^{\beta_1\cdots \beta_r}u'_{\beta_1}\cdots u'_{\beta_r}\ccr
&=ig\frac{i\varepsilon\cdot (p-p')}{(p^2+m^2)(p'^2+m^2)}(T_{(r)}^{a})_{\alpha_1\dots\alpha_r}^{\beta_1\cdots \beta_r}\, \frac{\bar u^{\alpha_1}\cdots \bar u^{\alpha_r} u'_{\beta_1}\cdots u'_{\beta_r}}{r!} ~,
\end{align}
that---after truncating the external scalar lines and stripping off the factor $\frac{\bar u\cdots u'}{r!}$---correctly reproduces the one-gluon two-scalar vertex
\begin{align}
\widetilde{\cal A}_{2s,1g}(p,\alpha;p',\beta;\varepsilon,k,a)= ig\, i\varepsilon\cdot (p-p') (T^a_{(r)})_{\alpha_1\dots \alpha_r}^{\beta_1\dots \beta_r}~.
\label{eq:1g2s-tr}
\end{align}

For the case of two external gluons the auxiliary field correlator involves two generators, and the color part reads
\begin{align}
&F^{a_1a_2}(\bar u,u')=\oint \frac{dz}{-2\pi i z}~\frac1{z^r}\, e^{\bar u\cdot u' z} \Big\langle \left({\bar u^{\alpha_1}}{z}+\bar \kappa^{\alpha_1}(\tau_1)\right)(T^{a_1})_{\alpha_1}{}^{\beta_1}\big( u_{\beta_1}+ \kappa_{\beta_1}(\tau_1)\big)\ccr &\hskip3cm \left({\bar u^{\alpha_2}}{z}+\bar \kappa^{\alpha_2}(\tau_2)\right)(T^{a_2})_{\alpha_2}{}^{\beta_2}\big( u_{\beta_2}+ \kappa_{\beta_2}(\tau_2)\big)\Big\rangle \ccr
&=\oint \frac{dz}{-2\pi i z}~e^{\bar u\cdot u' z}\Biggl( \frac1{z^{r-2}}\,\bar u T^{a_1} u'~\bar u T^{a_2} u'  +\frac1{z^{r-1}}\,\bar u T^{a_1} T^{a_2} u' ~\theta(\tau_1-\tau_2)+\ccr
&\hskip3cm +\frac1{z^{r-1}}\,\bar u T^{a_2} T^{a_1} u' ~\theta(\tau_2-\tau_1)+\frac1z\, {\rm tr}(T^{a_1} T^{a_2})\theta(\tau_1-\tau_2)\theta(\tau_2-\tau_1)\Biggr)~.
\end{align} 
The last term obviously vanishes for all $r$'s, whereas the first-one is non-zero when $r\geq 2$. Furthermore, by introducing $1=\theta(\tau_1-\tau_2)+\theta(\tau_2-\tau_1)$ in front of the first term, the latter reduces to
\begin{align}
& F^{a_1a_2}(\bar u,u')=\ccr
&\frac{1}{(r-1)!}\Big[\delta_{r\geq 2} (r-1) (\bar u u')^{r-2} \bar u T^{a_1} u'~\bar u T^{a_2} u' +  (\bar u u')^{r-1}\bar u T^{a_1} T^{a_2} u' \Big]\theta(\tau_1-\tau_2)\ccr 
&+\frac{1}{(r-1)!}\Big[\delta_{r\geq 2} (r-1) (\bar u u')^{r-2} \bar u T^{a_2} u'~\bar u T^{a_1} u' +  (\bar u u')^{r-1}\bar u T^{a_2} T^{a_1} u' \Big]\theta(\tau_2-\tau_1)~,
\end{align}
with an obvious notation for $\delta_{r\geq 2}$. The expressions in the square brackets are just the products of the generators in the rank-$r$ totally symmetric representation, decomposed in terms of the fundamental representation, i.e.
\begin{align}
 F^{a_1a_2}(\bar u,u')=\frac{\bar u^{\alpha_1}\cdots \bar u^{\alpha_r} u'_{\beta_1} \cdots u'_{\beta_r}}{r!}\Big[&\theta(\tau_1-\tau_2)  \big( T^{a_1}_{(r)} T^{a_2}_{(r)} \big)_{\alpha_1\dots \alpha_r}^{\beta_1\dots \beta_r}\ccr &  +\theta(\tau_2-\tau_1)  \big( T^{a_2}_{(r)} T^{a_1}_{(r)} \big)_{\alpha_1\dots \alpha_r}^{\beta_1\dots \beta_r} \Big]~,
\end{align} 
that, once again stripping off the prefactor $\frac{\bar u\cdots u'}{r!}$, gives the correct color factors for the two-gluon  term of the dressed scalar propagator. Finally, one can insert the latter into the formula~\eqref{eq:BK-QCD} and get the final result for the partial 
two-gluon scalar propagator: the integrals over $\tau_1$, $\tau_2$ and $T$ reproduce the correct form factors. In particular the term involving the expression $\varepsilon_1\cdot \varepsilon_2\, \delta(\tau_1-\tau_2)$ corresponds to the Feynman diagram with the two-gluon ``seagull'' vertex.

The  partial $n$-gluon scalar propagator term turns out to be a straightforward generalization of the previous expressions. In particular, for an even number of gluons, the color factor  displays the same features as the two-gluon one: One finds trace terms that  vanish as they multiply a full product of theta-functions. All the other terms combine into time-ordered products of color generators in the rank-$r$ totally symmetric representation, i.e.
\begin{align}
F^{a_1\dots a_n}(u,u') &=\frac{\bar u^{\alpha_1}\cdots \bar u^{\alpha_r} u'_{\beta_1} \cdots u'_{\beta_r}}{r!}\ccr&\times \sum_{\sigma\in S_n}\theta(\tau_{\sigma(1)}-\tau_{\sigma(2)})\cdots
\theta(\tau_{\sigma(n-1)}-\tau_{\sigma(n)}) \big( T^{a_{\sigma(1)}}_{(r)}\cdots T^{a_{\sigma(n)}}_{(r)} \big)_{\alpha_1\dots \alpha_r}^{\beta_1\dots \beta_r}~,
\end{align}    
where $S_n$ is the group of permutations of $n$ elements. The same happens for an odd number of gluons.

\subsection{Gauge-covariance of the dressed propagator and Ward identities}
The worldline representation for the dressed propagator in scalar QCD, eq.~\eqref{eq:WLgff}, comes in handy to show the gauge covariance of the propagator under finite transformations. 

Upon a finite gauge transformation $U(x)=e^{ig\epsilon(x)}$, with $\epsilon(x) = \epsilon^a(x) T^a$ and $(T^a)_\alpha{}^{\alpha'}$ in the fundamental representation, the gauge field transforms as
\begin{align}
\widetilde W_\mu(x) = U(x)\left[ \frac{i}{g}\partial_\mu +W_\mu(x) \right] U^\dagger(x)~,
\end{align}   
and the worldline lagrangian in~\eqref{eq:WLgff} is gauge-invariant provided the auxiliary fields transform as
\begin{align}
&c(\tau) \ \to  \ U(x(\tau)) c(\tau) \nonumber \\
&{\bar c} (\tau) \ \to \ {\bar c} (\tau) U^\dagger(x(\tau))~.
\label{eq:gauge-c}
\end{align} 
In particular, using that  $u'$ and $\bar u$ are the boundary values for $c(\tau)$ and $\bar c(\tau)$, at $\tau=0$ and $\tau=1$ respectively, the previous rule implies that
\begin{align}
& u'_\alpha \ \to \  (U(x'))_{\tilde\alpha}{}^{\alpha} u'_{\alpha} =:v'_{\tilde\alpha}\ccr
& \bar u^{\beta} \ \to \  \bar u^{\beta} (U^\dagger(x))_{\beta}{}^{\tilde \beta}=: \bar v^{\tilde\beta} ~,
\label{eq:gauge-u}
\end{align}
and
\begin{align}
\Big\langle \tilde \phi(x, {\bar v})\tilde{\bar \phi}(x', v')\Big\rangle_{\widetilde W}=\Big\langle \phi(x, \bar u){\bar \phi}(x', u')\Big\rangle_{W}
\end{align}
for the polarized scalar propagator. Using the transformation rules~\eqref{eq:gauge-u} and stripping off the color factors, we obtain
\begin{align}
&\Big\langle \tilde \phi_{\tilde \alpha_1\cdots \tilde \alpha_r}(x)\tilde{\bar \phi}^{\tilde\beta_1\cdots\tilde\beta_r }(x')\Big\rangle_{\widetilde W}\nonumber\\
&=(U(x))_{\tilde\alpha_1}{}^{\alpha_1}\cdots (U(x))_{\tilde\alpha_r}{}^{\alpha_r}\Big\langle \phi_{\alpha_1\cdots \alpha_r}(x){\bar \phi}^{\beta_1\cdots\beta_r }(x')\Big\rangle_{W}
(U^\dagger(x'))_{\beta_1}{}^{\tilde\beta_1}\cdots (U^\dagger(x'))_{\beta_r}{}^{\tilde\beta_r}~,
\end{align}
that is the correct transformation rule for the dressed propagator of a scalar field in the rank-$r$ totally symmetric representation of $SU(N)$. At the infinitesimal level, the previous expression yields a generating functional of Ward-identities for the above amplitudes, namely
\begin{align}
&0=D_\mu(W) \frac{\delta}{\delta W_\mu^a(y)} \Big\langle \phi_{\alpha_1\cdots \alpha_r}(x){\bar \phi}^{\beta_1\cdots\beta_r }(x')\Big\rangle_{W}\nonumber\\
&+ig\delta^{(D)}(y-x)\big(T^a_{(r)}\big)_{\alpha_1\cdots \alpha_r}^{\tilde\alpha_1\cdots \tilde\alpha_r}\Big\langle \phi_{\tilde\alpha_1\cdots \tilde\alpha_r}(x){\bar \phi}^{\beta_1\cdots\beta_r }(x')\Big\rangle_{W}\nonumber\\
&-ig\delta^{(D)}(y-x')\Big\langle \phi_{\alpha_1\cdots \alpha_r}(x){\bar \phi}^{\tilde\beta_1\cdots\tilde\beta_r }(x')\Big\rangle_{W}
\big(T^a_{(r)}\big)_{\tilde\beta_1\cdots \tilde\beta_r}^{\beta_1\cdots \beta_r}~.
\label{eq:WI-gen}
\end{align}
For example, at the leading order in the coupling constant $g$ the latter gives an identity between the one-gluon two-scalar term and the free scalar propagator. For definiteness let us consider $r=1$. In Fourier space, and using the truncated amplitudes, the previous functional identity reduces to the perturbative Ward identity 
\begin{align}
0=\delta^{(D)}(k+p+p') \Big[\widetilde {\cal A}_{2s,1g}(p,\alpha;p',\beta;-ik,k,a)+ig(T^a)_\alpha{}^\beta\big(p^2-p'^2\big) \Big]~,
\label{eq:WI-1g}
\end{align}
that does indeed vanish. Notice that in~\eqref{eq:WI-1g} the replacement $\varepsilon\to -ik$ implements the covariant derivative of the first term of~\eqref{eq:WI-gen}. 

\section{Conclusions and outlook}
\label{sec:concl}
We presented a worldline model where a scalar relativistic particle is coupled to a non-abelian gauge field. The quantization of the model yields the scalar propagator dressed by the external gauge field.
The path ordering---needed for the gauge-covariance of the model---is realized using (commuting) auxiliary fields that are coupled to the gauge potential. The addition of a worldline abelian gauge field along with a Chern-Simons term allows one to describe scalar particles that sit in an arbitrary rank-$r$ symmetric representation of the gauge group. We have specialized the non-abelian gauge field to be given by a sum of plane waves, and obtained 
explicit  expressions for the scalar propagator dressed by an arbitrary number $n$ of external gluons directly attached to the scalar line, the partial $n$-gluon scalar propagator. This is, 
in general, a gauge-dependent object, as three- and four-gluon vertices are not included,
but it is valid  off-shell  and it could be used as a building block for constructing
amplitudes. 
The resulting expressions are factorized into a color part and a kinematic part: their integral representation, provided by the present worldline model, may nicely combine with integration-by-parts techniques, that were shown to simplify the calculation of QCD form factors~\cite{Ahmadiniaz:2012xp, Ahmadiniaz:2012ie, Ahmadiniaz:2013rla}. 

One future line of research for extending our work would  be to find a simple way 
of adding the gluon self-interactions and  obtain the complete  tree-level amplitudes with 2 scalars and $n$ gluons.
One option would be to investigate a kind of  ``tree replacement rules'' of the type used successfully 
in similar Bern-Kosower formulas, where the scalar particle performs a loop 
instead of a line.

Also, the present approach can be generalized straightforwardly 
to scalar QCD, with scalar particles in a rank-$r$ totally antisymmetric representation, by using anticommuting auxiliary fields. Another quite welcome extension would be 
the treatment of the quark propagator in QCD, that may be obtained by coupling the external non-abelian field to a spinning particle, i.e. by using a particle with worldline supersymmetry, where the quantization of the spinorial coordinates describe the spin degrees of freedom. An alternative approach, that needs path ordering, is the inclusion of a matrix-valued (spin-factor) potential---see e.g. refs.~\cite{Feynman:1951gn,Bastianelli:1992ct,Gitman:1996wm}.

A possible application of the $n$-gluon master formula obtained above, would be the reconstruction, in any covariant gauge, of the scalar-gluon vertex  which is known as the Ball-Chiu vertex~\cite{ball-chiu-80}. 
In previous studies of form factors for QED and QCD---see refs.~\cite{Ahmadiniaz:2015new,Ahmadiniaz:2012xp,Ahmadiniaz:2013rla}---it was shown that the worldline formalism simplifies such calculations, and the results can be expressed in a much more compact way.  Finally it would be nice to extend the present treatment  by adding  the coupling to external gravity, 
along the lines of what has been done for the one-loop effective actions in curved space~\cite{Bastianelli:2002fv, Bastianelli:2002qw, Bastianelli:2005vk}.

\paragraph{Acknowledgments}{The authors thank Idrish Huet and Christian Schubert for helpful discussions. N.A. was supported by the PROMEP grant DSA/103.5/14/11184; he thanks to the Universit\`a di Bologna and the Universit\`a di Modena e Reggio Emilia for their warm hospitality while parts of this work were completed. O.C. was partly supported by the UCMEXUS--CONACYT grant CN-12-564; he thanks to the Departamento de F\'isica of the Universidad Nacional de La Plata for its warm hospitality while parts of this work were completed.}

\appendix   
\section{Coherent state path integral normalization}
\label{sec:app}
Let us consider a Hilbert space spanned by a complete set of harmonic oscillator states $\{\vert n\rangle\}$ with $n=0,1,\cdots,\infty$, and let $\hat c^\dagger$ and $\hat c$ be a pair of creation and annihilation operators acting on that Hilbert space and satisfying the canonical commutation relation $[\hat c, \hat c^\dagger]=1$.
These operators generate the harmonic oscillators states $\{\vert n\rangle\}$ in the usual
way.
Let us now define the ket and bra {\em coherent states} as right and left eigenstates of $\hat c$ and $\hat c^\dagger$,
respectively
\begin{align}
\hat c\vert u\rangle&= u\vert u\rangle~, \quad  \langle\bar u\vert \hat c^\dagger= \langle\bar u\vert \bar u~,
\end{align}
where $u$ and $\bar{u}$ are complex numbers. They can be constructed out of the Fock vacuum as 
\begin{align}
\vert u\rangle=e^{u \hat c^\dagger}\vert 0\rangle~,\quad\langle\bar u\vert=\langle0\vert e^{\bar{u} \hat c}~,
\end{align} 
and they form an overcomplete basis of the Hilbert space. Indeed their overlap reads
\begin{align}
\langle \bar u\vert u'\rangle=e^{\bar{u}u'}~.
\label{eq:zz'}
\end{align} 
After this introduction to bosonic coherent states we  prove the following identity: 
\begin{align}
\langle \bar{u}\vert e^{-i\theta \hat c^\dagger \hat c}\vert u'\rangle=e^{\bar{u}u'e^{-i\theta}}~.
\label{main-eq}
\end{align}
A simple way of proving it is to consider the wave function $\langle \bar u\vert \psi\rangle $, 
corresponding to a generic state $\vert \psi\rangle $ 
of the Hilbert space. On this wave function the creation and annihilation operators act as $\hat c^\dagger \to \bar u$
and $\hat c \to \frac{\partial}{\partial \bar u}$, so that the operator 
$ e^{-i\theta \hat c^\dagger \hat c} \to  e^{-i\theta  \bar u \frac{\partial}{\partial \bar u}}$
is just seen to generate a finite scaling of the $\bar u$ coordinate, namely $\bar u \to e^{-i\theta} \bar u $.
Then \eqref{main-eq} follows immediately. An alternative proof goes as follows.
Let us start from the ``normal order'' rule
\begin{align}
(\hat c^\dagger \hat c)^m=\sum_{k=0}^m S(m,k)(\hat c^\dagger)^k (\hat c)^k~,
\label{s-number}
\end{align}
where $S(m,k)$ are  the so-called ``Stirling numbers of second kind'' that are defined by~\footnote{These numbers 
appears often in reordering problems, as for example in the worldline two-loop computation of the Euler-Heisenberg effective lagrangian of scalar QED and spinor QED~\cite{Dunne:2002qg}.}
\begin{align}
S(m,k)=\frac{1}{k!}\sum_{l=0}^k\, (-)^l
\binom{k}{l}
(k-l)^m~,
\end{align}
with $S(n,0)=\delta_{n,0}$ and $S(n,1)=1$. Now from the definition and properties of coherent state and Eq. (\ref{s-number}) one can write the left hand side of Eq. (\ref{main-eq}) as 
\begin{align}
\langle \bar{u}\vert e^{-i\theta \hat c^\dagger \hat c}\vert u'\rangle &= \langle \bar{u}\vert \sum_{n=0}^\infty \frac{(-i\theta)^n}{n!}(a^\dagger a)^n\vert u'\rangle\nonumber\\
&=e^{\bar{u}u'}\sum_{n=0}^\infty \frac{(-i\theta)^n}{n!}\sum_{k=0}^n S(n,k) (\bar{u}u')^k
\nonumber \\
&= e^{\bar{u}u'}\sum_{n=0}^\infty \frac{(-i\theta)^n}{n!}T_n(\bar{u}u')~,
\label{lfhs}
\end{align}
where $T_n(x)$ are the so-called {\it Touchard} polynomials. Such polynomials can be also obtained through the exponential generating function 
\begin{align}
e^{x(e^{z}-1)}=\sum_{n=0}^\infty T_n(x)\frac{z^n}{n!}~.
\label{e4}
\end{align} 
Therefore, the series in~\eqref{lfhs} yields the right hand side of Eq.~(\ref{main-eq}), which is thus proved.

Expression~\eqref{main-eq} can thus be used to fix the normalization for the harmonic oscillator coherent state path integral. A free coherent state path integral can be built up by inserting spectral decompositions of the identity
\begin{align}
{\mathbb 1} = \int \frac{d\bar c\, d c}{2\pi i}~e^{-\bar c c}|c\rangle \langle \bar c|  
\end{align}
into the expression~\eqref{eq:zz'} for the scalar product. One gets
\begin{align}
\langle \bar u| u'\rangle = e^{\bar u u'} =\int \prod_{k=1}^{n-1}\frac{d\bar c_k dc_k}{2\pi i}~\exp\left\{\bar c_n c_{n-1} -\sum_{k=1}^{n-1}\bar c_k (c_k-c_{k-1})\right\} ~,
\end{align}  
where we defined $c_0:=u'$ and $\bar c_n := \bar u$. Furthermore we define a time parameter $0\leq \tau \leq 1$ and identify $c(\tau_k):=c_k$ and $\bar c(\tau_k):= \bar c_k$, with $\tau_k-\tau_{k-1}=\frac{1}{n-1}=:\epsilon$. Thus, in the large $n$ limit we may identify the latter as the free coherent state path integral
\begin{align}
\langle \bar u| u'\rangle = e^{\bar u u'}  = \int_{c(0)=u'}^{\bar c(1)=\bar u} D\bar c Dc ~e^{-S_f[c,\bar c]}~,
\end{align}
with
\begin{align}
S_f[c,\bar c] = \int_0^1d\tau ~\bar c(\tau) \dot c(\tau) -\bar c c(1)~. 
\end{align}
We may thus split the paths  into backgrounds, satisfying the free equations of motion $\dot {\bar c}=\dot c =0$ with corresponding boundary conditions, and quantum fluctuations
   \begin{align}
&c(\tau) = u' +\kappa(\tau)\,,\quad \kappa(0) =0 \\
&\bar c(\tau) = \bar u +\bar \kappa(\tau)\,,\quad \bar\kappa(1) =0~.
\end{align} 
Hence,
\begin{align}
\langle \bar u| u'\rangle = e^{\bar u u'}  = e^{\bar u u'}\int_{\kappa(0)=0}^{\bar \kappa(1)=0} D\bar \kappa D\kappa ~e^{-\int_0^1d\tau \bar\kappa \dot \kappa}~,
\end{align}
that in turn gives
\begin{align}
\int_{\kappa(0)=0}^{\bar \kappa(1)=0} D\bar \kappa D\kappa ~e^{-\int_0^1d\tau \bar\kappa \dot \kappa} =1~.
\label{eq:Z1}
\end{align}
Let us now consider the path integral
\begin{align}
 Z(\bar u,u';\theta):=\int_{c(0)=u'}^{\bar c(1)=\bar u} D\bar c Dc ~e^{-S[c,\bar c;\theta]}~,
\end{align}
with
\begin{align}
S[c,\bar c] = \int_0^1d\tau ~\bar c(\tau) \big(\partial_\tau+i\theta\big) c(\tau) -\bar c c(1)~.
\end{align}
We can solve the latter in a similar way as above, i.e. by splitting the paths into backgrounds 
\begin{align}
&\big(\partial_\tau+i\theta\big) c(\tau)=0\,,\ c(0)=u'\quad\Rightarrow\quad C(\tau) = u'~e^{-i\theta\tau}\\
&\big(-\partial_\tau+i\theta\big) \bar c(\tau)=0\,,\ \bar c(1)=\bar u\quad\Rightarrow\quad \bar C(\tau) = \bar u~e^{i\theta(\tau-1)}~,
\end{align} 
and quantum fluctuations $\kappa(\tau)$ and $\bar\kappa(\tau)$ with the same boundary conditions as above. By setting
\begin{align}
&c(\tau) = e^{-i\theta\tau} (u' + \kappa(\tau))\\
&\bar c(\tau) =e^{i\theta\tau}(e^{-i\theta}\bar u+ \bar \kappa(\tau) )~,
\end{align}
one gets
\begin{align}
Z(\bar u,u';\theta):=e^{e^{-i\theta}\bar u u'}\int_{\kappa(0)=0}^{\bar \kappa(1)=\bar 0} D\bar \kappa D\kappa ~e^{-\int_0^1d\tau \bar\kappa \dot \kappa} =e^{e^{-i\theta}\bar u u'}~,
\end{align} 
where we made use of~\eqref{eq:Z1}. In summary,
\begin{align}
\int_{c(0)=u'}^{\bar c(1)=\bar u} D\bar c Dc ~e^{-\int_0^1d\tau\, \bar c(\tau) \big( \partial_\tau+i\theta \big) c(\tau) +\bar c c(1)}=e^{e^{-i\theta}\bar u u'}~,
\label{eq:pathN}
\end{align}
which we thus identify with~\eqref{main-eq}. Thus, by adding a constant term to the action and considering $\alpha=1,\dots,N$ independent pairs of oscillator operators, we finally get 
\begin{align}
\int_{c(0)=u'}^{\bar c(1)=\bar u} D\bar c Dc ~e^{-\int_0^1d\tau\, \bar c^\alpha(\tau)\big( \partial_\tau+i\theta \big) c_\alpha(\tau) +\bar c \cdot c(1)+i\theta r} &= \langle \bar u\vert e^{-i\theta \big(\hat c^\dagger \cdot \hat c -r\big)} \vert u'\rangle\nonumber\\
&=e^{e^{-i\theta}\bar u\cdot u' +i\theta r}~,
\label{eq:path-normal}
\end{align} 
where ${\mathbb N}:= \hat c^\dagger \cdot \hat c =\hat c^\dagger{}^\alpha\, \hat c_\alpha$ is the total occupation number operator for the system of $N$ harmonic oscillators.

If, on the other hand, one were to identify
\begin{align}
\int_{c(0)=u'}^{\bar c(1)=\bar u} D\bar c Dc ~e^{-\int_0^1d\tau\, \bar c^\alpha(\tau)\big( \partial_\tau+i\theta \big) c_\alpha(\tau) +\bar c \cdot c(1)} = \langle \bar u\vert e^{-i\theta (\hat c^\dagger \cdot \hat c)_s} \vert u'\rangle ~,
\end{align}
with 
\begin{align}
(\hat c^\dagger \cdot \hat c)_s := \frac{1}{2}\big(\hat c^\dagger \cdot \hat c +\hat c \cdot \hat c^\dagger\big) =\hat c^\dagger \cdot \hat c +\frac{N}{2}
\end{align}
being the symmetrized product, one would then get 
\begin{align}
\int_{c(0)=u'}^{\bar c(1)=\bar u} D\bar c Dc ~e^{-\int_0^1d\tau\, \bar c^\alpha(\tau)\big( \partial_\tau+i\theta \big) c_\alpha(\tau) +\bar c \cdot c(1)} =e^{e^{-i\theta}\bar u\cdot u'-i\theta\frac{N}{2}} ~,
\end{align}
and
\begin{align}
\int_{c(0)=u'}^{\bar c(1)=\bar u} D\bar c Dc ~e^{-\int_0^1d\tau\, \bar c^\alpha(\tau)\big( \partial_\tau+i\theta \big) c_\alpha(\tau) +\bar c \cdot c(1)+i\theta s} &= \langle \bar u\vert e^{-i\theta \big((\hat c^\dagger \cdot \hat c)_s -s\big)} \vert u'\rangle\nonumber\\
&=e^{e^{-i\theta}\bar u\cdot u'+i\theta(s-\frac{N}{2})} ~.
\label{eq:path-symmetric}
\end{align}
However, setting the eigenvalues of the occupation numbers to be equal
\begin{align}
(\hat c^\dagger \cdot \hat c)_s -s= \hat c^\dagger \cdot \hat c -r\quad \Longrightarrow\quad s=r+\frac{N}{2}~,
\end{align} 
the right hand sides of~\eqref{eq:path-normal} and~\eqref{eq:path-symmetric} coincide.



\begin{thebibliography}{99}

\bibitem{Strassler:1992zr}
  M.~J.~Strassler,
  ``Field theory without Feynman diagrams: one loop effective actions,''
  Nucl.\ Phys.\ B {\bf 385} (1992) 145
  [hep-ph/9205205].
  
\bibitem{Bern:1990cu}
  Z.~Bern and D.~A.~Kosower,
  ``Efficient calculation of one loop QCD amplitudes,''
  Phys.\ Rev.\ Lett.\  {\bf 66} (1991) 1669.
  
\bibitem{Bern:1991aq}
  Z.~Bern and D.~A.~Kosower,
  ``The computation of loop amplitudes in gauge theories,''
  Nucl.\ Phys.\ B {\bf 379} (1992) 451.
  
\bibitem{Schubert:2001he}
  C.~Schubert,
  ``Perturbative quantum field theory in the string inspired formalism,''
  Phys.\ Rept.\  {\bf 355} (2001) 73
  [hep-th/0101036].
  
\bibitem{Schmidt:1994aq}
  M.~G.~Schmidt and C.~Schubert,
  ``Multiloop calculations in the string inspired formalism: the single spinor loop in QED,''
  Phys.\ Rev.\ D {\bf 53} (1996) 2150
  [hep-th/9410100].
  
\bibitem{Reuter:1996zm}
  M.~Reuter, M.~G.~Schmidt and C.~Schubert,
  ``Constant external fields in gauge theory and the spin 0, 1/2, 1 path integrals,''
  Annals Phys.\  {\bf 259} (1997) 313
  [hep-th/9610191].\\
  
\bibitem{Gies:2005sb}
  H.~Gies, J.~Sanchez-Guillen and R.~A.~Vazquez,
  ``Quantum effective actions from nonperturbative worldline dynamics,''
  JHEP {\bf 0508} (2005) 067
  [hep-th/0505275].
  
\bibitem{Dunne:2005sx}
  G.~V.~Dunne and C.~Schubert,
  ``Worldline instantons and pair production in inhomogeneous fields,''
  Phys.\ Rev.\ D {\bf 72} (2005) 105004
  [hep-th/0507174].
  
\bibitem{Gies:2003cv}
  H.~Gies, K.~Langfeld and L.~Moyaerts,
  ``Casimir effect on the worldline,''
  JHEP {\bf 0306} (2003) 018
  [hep-th/0303264].
 
\bibitem{Bastianelli:2002fv}
  F.~Bastianelli and A.~Zirotti,
  ``Worldline formalism in a gravitational background,''
  Nucl.\ Phys.\ B {\bf 642} (2002) 372
  [hep-th/0205182].
  
\bibitem{Bastianelli:2002qw}
  F.~Bastianelli, O.~Corradini and A.~Zirotti,
  ``Dimensional regularization for N=1 supersymmetric sigma models and the worldline formalism,''
  Phys.\ Rev.\ D {\bf 67} (2003) 104009
  [hep-th/0211134].
  
\bibitem{Bastianelli:2005vk}
  F.~Bastianelli, P.~Benincasa and S.~Giombi,
  ``Worldline approach to vector and antisymmetric tensor fields,''
  JHEP {\bf 0504} (2005) 010
  [hep-th/0503155].

\bibitem{Bastianelli:2004zp}
  F.~Bastianelli and C.~Schubert,
  ``One loop photon-graviton mixing in an electromagnetic field: Part 1,''
  JHEP {\bf 0502} (2005) 069
  [gr-qc/0412095].
    
\bibitem{Bastianelli:2007pv}
  F.~Bastianelli, O.~Corradini and E.~Latini,
  ``Higher spin fields from a worldline perspective,''
  JHEP {\bf 0702} (2007) 072
  [hep-th/0701055].
  
\bibitem{Bastianelli:2008nm}
  F.~Bastianelli, O.~Corradini and E.~Latini,
  ``Spinning particles and higher spin fields on (A)dS backgrounds,''
  JHEP {\bf 0811} (2008) 054
  [arXiv:0810.0188 [hep-th]].
  
\bibitem{Bastianelli:2012bn}
  F.~Bastianelli, R.~Bonezzi, O.~Corradini and E.~Latini,
  ``Effective action for higher spin fields on (A)dS backgrounds,''
  JHEP {\bf 1212} (2012) 113
  [arXiv:1210.4649 [hep-th]].
  
\bibitem{Dai:2008bh}
  P.~Dai, Y.~-t.~Huang and W.~Siegel,
  ``Worldgraph approach to Yang-Mills amplitudes from N=2 spinning particle,''
  JHEP {\bf 0810} (2008) 027
  [arXiv:0807.0391 [hep-th]].
  
\bibitem{Fliegner:1993wh}
  D.~Fliegner, M.~G.~Schmidt and C.~Schubert,
  ``The higher derivative expansion of the effective action by the string inspired method. Part 1.,''
  Z.\ Phys.\ C {\bf 64} (1994) 111
  [hep-ph/9401221].

\bibitem{Fliegner:1997rk}
  D.~Fliegner, P.~Haberl, M.~G.~Schmidt and C.~Schubert,
  ``The higher derivative expansion of the effective action by the string inspired method. Part 2,''
  Annals Phys.\  {\bf 264} (1998) 51
  [hep-th/9707189].
  
\bibitem{Bastianelli:2008vh}
  F.~Bastianelli, O.~Corradini, P.~A.~G.~Pisani and C.~Schubert,
  ``Scalar heat kernel with boundary in the worldline formalism,''
  JHEP {\bf 0810} (2008) 095
  [arXiv:0809.0652 [hep-th]].
  
\bibitem{Kiem:2001dk}
  Y.~Kiem, S.~J.~Rey, H.~T.~Sato and J.~T.~Yee,
  ``Anatomy of one loop effective action in noncommutative scalar field theories,''
  Eur.\ Phys.\ J.\ C {\bf 22} (2002) 757
  [hep-th/0107106].
  
\bibitem{Bonezzi:2012vr}
  R.~Bonezzi, O.~Corradini, S.~A.~Franchino Vinas and P.~A.~G.~Pisani,
  ``Worldline approach to noncommutative field theory,''
  J.\ Phys.\ A {\bf 45} (2012) 405401
  [arXiv:1204.1013 [hep-th]].
  
\bibitem{Mansfield:2014vea}
  P.~Mansfield,
  ``The fermion content of the Standard Model from a simple world-line theory,''
  Phys.\ Lett.\ B {\bf 743} (2015) 353
  [arXiv:1410.7298 [hep-ph]].
  
\bibitem{Edwards:2014bfa}
  J.~P.~Edwards,
  ``Unified theory in the worldline approach,''
  arXiv:1411.6540 [hep-th].
  
\bibitem{Feynman:1951gn}
  R.~P.~Feynman,
  ``An operator calculus having applications in quantum electrodynamics,''
  Phys.\ Rev.\  {\bf 84} (1951) 108.
  
    
\bibitem{Daikouji:1995dz}
  K.~Daikouji, M.~Shino and Y.~Sumino,
  ``Bern-Kosower rule for scalar QED,''
  Phys.\ Rev.\ D {\bf 53} (1996) 4598
  [hep-ph/9508377].

\bibitem{Bastianelli:2014bfa}
  F.~Bastianelli, A.~Huet, C.~Schubert, R.~Thakur and A.~Weber,
  ``Integral representations combining ladders and crossed-ladders,''
  JHEP {\bf 1407} (2014) 066
  [arXiv:1405.7770 [hep-ph]].
  
\bibitem{Bern:2008qj}
  Z.~Bern, J.~J.~M.~Carrasco and H.~Johansson,
  ``New relations for gauge-theory amplitudes,''
  Phys.\ Rev.\ D {\bf 78} (2008) 085011
  [arXiv:0805.3993 [hep-ph]].
  
\bibitem{Fradkin:1955jr}
  E.~S.~Fradkin,
  ``Concerning some general relations of quantum electrodynamics,''
  Zh.\ Eksp.\ Teor.\ Fiz.\  {\bf 29} (1955) 258
   [Sov.\ Phys.\ JETP {\bf 2} (1956) 361].

\bibitem{Landau:1955zz}
  L.~D.~Landau and I.~M.~Khalatnikov,
  ``The gauge transformation of the Green function for charged particles,''
  Sov.\ Phys.\ JETP {\bf 2} (1956) 69
   [Zh.\ Eksp.\ Teor.\ Fiz.\  {\bf 29} (1955) 89].
  
  \bibitem{Ahmadiniaz:2015new}
  N. Ahmadiniaz, A. Bashir and C. Schubert, in preparation.
  
\bibitem{Badger:2005zh}
  S.~D.~Badger, E.~W.~N.~Glover, V.~V.~Khoze and P.~Svrcek,
  ``Recursion relations for gauge theory amplitudes with massive particles,''
  JHEP {\bf 0507} (2005) 025
  [hep-th/0504159].
  
\bibitem{Forde:2005ue}
  D.~Forde and D.~A.~Kosower,
  ``All-multiplicity amplitudes with massive scalars,''
  Phys.\ Rev.\ D {\bf 73} (2006) 065007
  [hep-th/0507292].
  
\bibitem{Ferrario:2006np}
  P.~Ferrario, G.~Rodrigo and P.~Talavera,
  ``Compact multigluonic scattering amplitudes with heavy scalars and fermions,''
  Phys.\ Rev.\ Lett.\  {\bf 96} (2006) 182001
  [hep-th/0602043].
  
\bibitem{Boels:2007pj}
  R.~Boels and C.~Schwinn,
  ``CSW rules for a massive scalar,''
  Phys.\ Lett.\ B {\bf 662} (2008) 80
  [arXiv:0712.3409 [hep-th]].
  
\bibitem{Bastianelli:2013pta}
  F.~Bastianelli, R.~Bonezzi, O.~Corradini and E.~Latini,
  ``Particles with non abelian charges,''
  JHEP {\bf 1310} (2013) 098
  [arXiv:1309.1608 [hep-th]].
  
\bibitem{Bastianelli:2015iba}
  F.~Bastianelli, R.~Bonezzi, O.~Corradini, E.~Latini and K.~H.~Ould-Lahoucine,
  ``A worldline approach to colored particles,''
  arXiv:1504.03617 [hep-th].
  
\bibitem{Balachandran:1976ya}
  A.~P.~Balachandran, P.~Salomonson, B.~S.~Skagerstam and J.~O.~Winnberg,
  ``Classical description of particle interacting with nonabelian gauge field,''
  Phys.\ Rev.\ D {\bf 15} (1977) 2308.

\bibitem{Barducci:1976xq}
  A.~Barducci, R.~Casalbuoni and L.~Lusanna,
  ``Classical scalar and spinning particles interacting with external Yang-Mills fields,''
  Nucl.\ Phys.\ B {\bf 124} (1977) 93.
  
\bibitem{D'Hoker:1995bj}
  E.~D'Hoker and D.~G.~Gagne,
  ``Worldline path integrals for fermions with general couplings,''
  Nucl.\ Phys.\ B {\bf 467} (1996) 297
  [hep-th/9512080].
  
\bibitem{Fradkin:1966zz}
  E.~Fradkin,
  ``Application of functional methods in quantum field theory and quantum statistics (II),''
  Nucl.\ Phys.\  {\bf 76} (1966) 588.

\bibitem{Howe:1988ft}
  P.~S.~Howe, S.~Penati, M.~Pernici and P.~K.~Townsend,
  ``Wave equations for arbitrary spin from quantization of the extended supersymmetric spinning particle,''
  Phys.\ Lett.\ B {\bf 215} (1988) 555.
  
\bibitem{Ahmadiniaz:2012xp}
  N.~Ahmadiniaz and C.~Schubert,
  ``A covariant representation of the Ball-Chiu vertex,''
  Nucl.\ Phys.\ B {\bf 869} (2013) 417
  [arXiv:1210.2331 [hep-ph]].
   
\bibitem{Ahmadiniaz:2012ie}
  N.~Ahmadiniaz, C.~Schubert and V.~M.~Villanueva,
  ``String-inspired representations of photon/gluon amplitudes,''
  JHEP {\bf 1301} (2013) 132
  [arXiv:1211.1821 [hep-th]].
  
\bibitem{Ahmadiniaz:2013rla}
  N.~Ahmadiniaz and C.~Schubert,
  ``Form factor decomposition of the off-shell four-gluon amplitudes,''
  PoS QCD {\bf -TNT-III} (2013) 002
  [arXiv:1311.6829 [hep-ph]].
  
\bibitem{Bastianelli:1992ct}
  F.~Bastianelli and P.~van Nieuwenhuizen,
  ``Trace anomalies from quantum mechanics,''
  Nucl.\ Phys.\ B {\bf 389} (1993) 53
  [hep-th/9208059].
  
\bibitem{Gitman:1996wm}
  D.~M.~Gitman and S.~I.~Zlatev,
  ``Spin factor in path integral representation for Dirac propagator in external fields,''
  Phys.\ Rev.\ D {\bf 55} (1997) 7701
  [hep-th/9608179].

\bibitem{Dunne:2002qg}
  G.~V.~Dunne and C.~Schubert,
  ``Two loop selfdual Euler-Heisenberg Lagrangians. 2. Imaginary part and Borel analysis,''
  JHEP {\bf 0206} (2002) 042
  [hep-th/0205005].
  
  \bibitem{ball-chiu-80}
  J. S. Ball and T-W. Chiu, 
  ``Analytic properties of the vertex function in gauge theories. II''
    Phys.\ Rev.\ D {\bf 22} (1980) 2550; Erratum ibdio {\bf23} (1981) 3085.

\end{thebibliography}
\end{document}